\documentstyle[aps,twocolumn,prb]{revtex}
\begin{document}
\draft

\title{Monte Carlo Simulation of Magnetization Reversal in Fe 
       Sesquilayers on W(110)}

\author{M. Kolesik}
\address{
Supercomputer Computations Research Institute, 
Florida State University, Tallahassee, Florida 32306-4052 \\
and 
Institute of Physics, Slovak Academy of Sciences,
D\' ubravsk\' a cesta 9, 84228 Bratislava, Slovak Republic
        }

\author{M. A. Novotny}
\address{  
Supercomputer Computations Research Institute, 
Florida State University, Tallahassee, Florida 32306-4052 \\
and
Department of Electrical Engineering,
2525 Pottsdamer Street,   
Florida A\&M University--Florida State University,
Tallahassee, Florida 32310-6046
        }

\author{Per Arne Rikvold}
\address{  
Center for Materials Research and Technology,
Department of Physics, \\
and 
Supercomputer Computations Research Institute, 
Florida State University, Tallahassee, Florida 32306-3016 \\
        }

\date{\today}
\maketitle
\begin{abstract}

Iron sesquilayers grown at room temperature on W(110)  exhibit a pronounced
coercivity maximum near a coverage of 1.5 atomic monolayers.  On lattices
which faithfully reproduce the morphology of the real films, a kinetic Ising
model is utilized to simulate the domain-wall motion.  Simulations reveal
that the dynamics is dominated by the second-layer islands, which act as
pinning centers. The simulated dependencies of the coercivity on the film
coverage, as well as on the temperature and the frequency of the applied
field, are very similar to those measured in experiments.  Unlike previous
micromagnetic models, the presented approach provides insight into the
dynamics of the domain-wall motion and clearly reveals the role of thermal
fluctuations. 

\end{abstract}

\section{Introduction}

Recently, there has been much interest in ultrathin iron films on W(110) 
substrates. 
\cite{Bethge95,Sander96a,Sander96b,Skomski96,Elmers95,Elmers94,Elmers90,Suen} 
The
present work is mainly concerned with the so-called sesquilayers, which are
films with coverages between one and two atomic monolayers. When grown at
room temperature, such structures consist of a nearly perfect monolayer with
compact islands of the second layer on top.\cite{Bethge95} The mechanical
and magnetic properties are profoundly affected by this
morphology.\cite{Sander96a,Sander96b,Skomski96} For example, the coercivity
exhibits a pronounced maximum as a function of the sesquilayer coverage. In
particular, around a coverage of 1.5 monolayers (ML) the coercivity exceeds
that of a monolayer or a doublelayer by more than an order of
magnitude.\cite{Skomski96} Similarly, elastic properties show an unusual
behavior in this thickness range.\cite{Sander96a,Sander96b} The reportedly
observed absence of magnetic long-range order in a certain region of
coverage was first interpreted as a manifestation of a spin-glass-like
phase.\cite{Elmers95} However, the required presence of frustrated
antiferromagnetic interactions was difficult to justify, and later an
explanation based on surface roughness was
proposed.\cite{Sander96a,Skomski96}

The aim of the present article is to study a model of this system by means
of computer simulation. Our goal is twofold.  First, we provide a
semiquantitative explanation for the experimental coercivity observations.
In spirit, this is similar to the micromagnetic models proposed in
Refs.~\onlinecite{Sander96a} and \onlinecite{Skomski96}, but in contrast to
the essentially zero-temperature micromagnetic arguments it gives an insight
into the dynamics of the magnetization reversal at nonzero temperatures. Our
model reproduces without additional assumptions the experimentally observed
temperature\cite{Sander96a} and frequency\cite{Suen} dependence of the
coercivity.  Second, we want to demonstrate the usefulness of the Monte
Carlo approach to the simulation of real magnetic systems. This is an
application-oriented extension of our previous work on magnetization
switching in kinetic Ising models,\cite{Rikvold,Richards} where one can find
the theoretical background for the present article. Some preliminary results
of this study have been discussed in Ref.~\onlinecite{Geilo97}. 

The remainder of this paper is organized as follows. In the next section we
describe how the dynamic magnetic properties of iron sesquilayers are
reflected by our computational model.  Section III is devoted to simulation
results and their comparison with experimental data. A summary of our
results, as well as a brief discussion of the implications of the proposed
model for the frequency dependence of the coercivity are given in Sec. IV. 

\section{Computational model of an ultrathin iron film}

From the theoretical point of view, an iron monolayer deposited on a W(110) 
substrate is a nearly ideal ferromagnetic Ising-like system.\cite{Elmers94}
Although this only implies that the {\em critical fluctuations} in the
vicinity of the Curie temperature can be described by the Ising model, one
is tempted to speculate that it could be used also away from the phase
transition. However, what is more important at lower temperatures is the
magnetic anisotropy. The magnetocrystalline anisotropy of the distorted bcc
iron structure on the tungsten substrate is almost a hundred times larger
than in bulk iron.\cite{Elmers90} The strength of the resulting anisotropy
field, on the order of a few T,\cite{Skomski96} makes the Ising model a
reasonable approximation for this system. The Ising Hamiltonian is given by
\begin{equation}
{\cal H} = -J \sum_{\langle ij \rangle } s_i s_j - \mu H \sum_{i} s_i \ .
\label{eq:ham}
\end{equation}
Here, the two-state variables $s_i = \pm 1$ at the nodes of a computational
lattice represent local magnetic moments pointing in the positive or
negative in-plane easy-axis [$\bar 1$10] directions, respectively. The first
sum runs over all nearest-neighbor pairs in the lattice. One and two layers
of a square lattice are used to model mono- and double-layer regions of the
film, respectively.  Therefore, the number of nearest neighbors for a given
site depends on its environment. The ferromagnetic spin-spin interaction,
$J=8.73$ meV, is fixed such that the critical temperature, $T_c=230$~K, of a
monolayer\cite{Elmers94} is correctly reproduced by the exact value for our
square computational lattice. The second term, with the summation over all
sites, represents the interaction with the component $H$ of the external
magnetic field along the easy axis, and $\mu$ is the magnetic moment of the
computational spin $s_i$. The latter depends on the model lattice spacing
$a$, which we choose to be $a=6~$\AA. With this value of $a$, a
computational spin $s_i$ represents a part of an iron monolayer containing
approximately 5 atoms and carries a magnetic moment of $\mu = 11.43~\mu_B$. 
We use the bulk value of the iron atom magnetic moment.

\begin{figure}[t]
\vspace{5.9in} 
\includegraphics{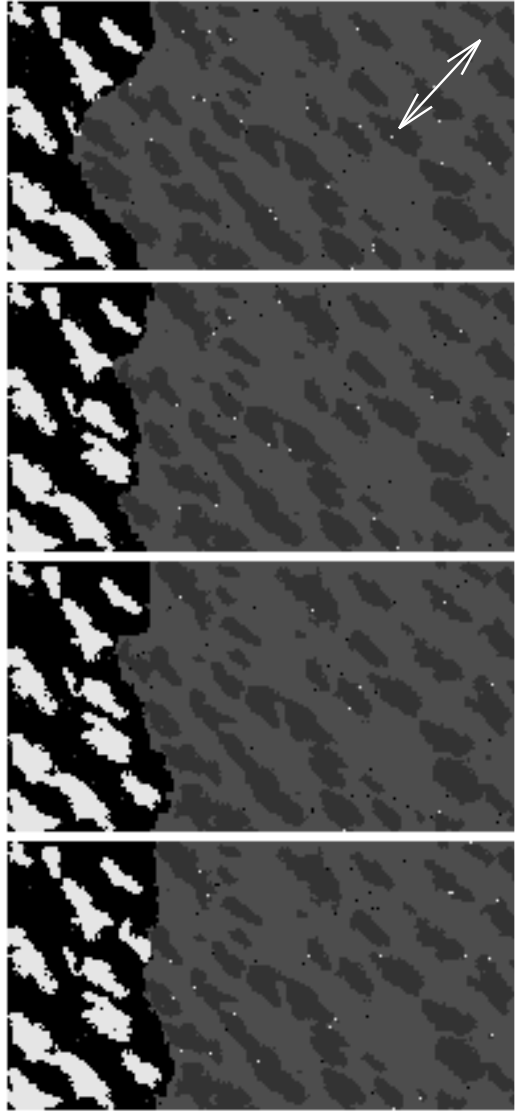} 
\caption[]{ Illustration of the sesquilayer morphology at a coverage of 1.26
ML, with snapshots of the propagating domain wall at the times marked by
arrows in Fig.~2. Time increases from top to bottom in the picture.  The
patches are the islands of the second iron monolayer on top of the first
one. The area shown is 1170 \AA $\times$ 610 \AA, and the island
configuration was digitized from Fig.1 j) of
Ref.~\protect{\onlinecite{Bethge95}}.  The size of the computational lattice
was $195\times102\times1(2)$ (two layers for islands). $T = 1.3\ J/k_{\rm B} 
\approx 132$~K, $\mu H=0.02\ J$ or $H\approx 0.26$\ T.  
The high-contrast region is the stable magnetic phase. 
The double arrow shows the easy-axis direction.
An animation of this simulation can be found at 
{\tt http://www.scri.fsu.edu/ $\widetilde{}\ $rikvold}.
\label{fig:snap}}  
\end{figure}

The morphology of the iron sesquilayers has been well studied by scanning
tunneling microscopy (STM), and high-resolution images of their surfaces are
available. Since the roughness of the surface plays a crucial role in the
magnetic behavior, we have chosen to digitize STM pictures from
Ref.~\onlinecite{Bethge95} to generate our computational lattices.  This has
the advantage that the lattices faithfully reproduce the real films. 
Figure~\ref{fig:snap} shows the morphology of the iron sesquilayers.

The dynamics we use in our simulations is the standard Glauber Monte Carlo
dynamics with local updates at randomly chosen sites, which can be related
to the quantum mechanical dynamics.\cite{Martin77} To speed up the
simulations in weak external fields, we use a variant of the rejection-free
Monte Carlo algorithm described in Ref.~\onlinecite{Novotny}.  The
simulation time is measured in Monte Carlo Steps per Spin (MCSS). Although
its precise relation to physical time is not known, it is expected that one
MCSS roughly corresponds to a typical inverse phonon frequency. Here, we fix
the simulation time scale such that 1 MCSS $\equiv t_{\rm MC}=10^{-12}$s. 

\newpage
\null

\newpage

Before describing how we estimate the coercivity from simulations, let us
consider the possible magnetization reversal modes in real systems.
There are two extremes for magnetization
switching:  driven either by nucleation or by domain-wall motion. What
is observed in experiments on these ultrathin iron films is the second
mechanism. On a microscopic scale, the sample is ``infinite,'' and somewhere
there exist seeds of the stable phase which start to grow almost
instantaneously when the field is reversed.  As a result, magnetization
reversal in a typical region of the sample is caused by a domain wall which
propagates across the observed area. When the driving magnetic field is weak
and/or the disorder is strong, the domain wall is pinned and does not move
on the experimental time scale. Thus, the coercivity can be identified with
the field which makes the domain wall mobile under given circumstances
(disorder, temperature, time scale, ...).  To exclude the nucleation time
from our simulations, we prepare the initial state with all spins equal to
$+1$, except for those in a narrow strip along one of the short sides of the
lattice. These spins are initialized to $-1$ and represent a domain of the
stable magnetization that has just propagated into the region under
observation. We apply a negative magnetic field and measure the time needed,
$t_{\rm sw}$, to reach the state with 90\% of the spins reversed. Since the reversal
occurs by propagation of the domain wall, the change in the the reversed
volume divided by the lattice width gives us the average distance $D$ the
domain wall traveled.  Together with the elapsed time it can be used to
estimate the average domain-wall velocity.

\section{Simulation results}

Figure~\ref{fig:snap} illustrates the propagation of a domain wall through a
sesquilayer system as it was recorded during a particular simulation run.
One notices that the domain wall typically does not cross the
second-monolayer patches.  Instead, it feels an island as an obstacle and
gets pinned near its boundary.  To overturn the spins in the island, it is
necessary to overcome a free-energy barrier which depends on the driving
field, the temperature, and also on the shape and size of the island.  Due
to thermal fluctuations, the domain wall tries to enter the islands, but is
usually driven back by the free-energy barrier. A ``successful'' fluctuation
must be large enough to have greater probability to grow through the island
than to shrink and vanish; a fluctuation which just meets this criterion is
called {\em critical}.

To estimate the size of a critical fluctuation, denote by $\ell$ the linear
extent of the region which reverses its magnetization during a jump of the
domain wall to its next metastable configuration. The interface energy
change $\delta E_\Sigma$ is proportional to $\ell$, while the ``volume''
energy change $\delta E_H$ is proportional to $H \ell^2$.  While the latter
is negative, the interface energy change can be negative or positive
depending on whether the domain-wall length decreases or increases.  The
mobility of the domain wall is essentially given by the slower processes,
for which $\delta E_\Sigma > 0$ and $\delta E_H < 0$.  The new configuration
will be stable with respect to the previous one only if the volume energy
change is sufficient to compensate for the interface energy change.  From
this it follows that $\ell \sim 1/H$. Since the interface and volume
energies of the intermediate unstable domain-wall configurations scale in
the same way, the free-energy barrier $\Delta F$ to be overcome will be
inversely proportional to the driving field, $\Delta F = \Xi/(\mu H)$. The
field-independent quantity $\Xi$ depends on the local environment, as well as
on the temperature.  In such a situation, nucleation theory predicts that
the waiting time needed to observe a critical fluctuation
is\cite{Rikvold,Richards}
\begin{equation}
\tau \sim H^{-K} \exp\left( {\Xi \over k_{\rm B} T \> \mu H }  \right) \ ,
\label{eq:nucl}
\end{equation}
where the exponent $K$ is taken to be equal to $3$ as expected for the
two-dimensional Ising model,\cite{Rikvold} and $k_{\rm B}$ is
Boltzmann's constant. Once a critical fluctuation is
created, the domain wall reorganizes quickly until it reaches a new
metastable configuration of lower free energy than the previous one. In this
way, propagation of a domain wall through the sesquilayer is explained as a
series of thermally induced nucleation events followed by rapid
magnetization changes.

\begin{figure}[t]
\vspace*{3in} 
\includegraphics{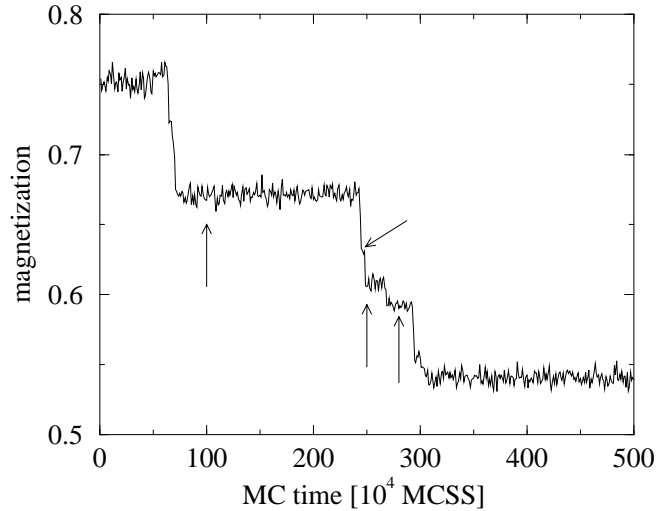} 
\caption[]{ Magnetization per spin vs. time during the same simulation from
which the snapshots in Fig. 1 were taken (at the times indicated by arrows). 
Most of the time, the magnetization remains constant, apart from small 
fluctuations. The movement of the domain wall occurs in jumps. Each jump 
represents a nucleation event, after which the domain wall advances quickly 
to its new metastable configuration.
\label{fig:magt}} 
\end{figure}

Figure~\ref{fig:magt} shows how the intermittent domain-wall motion is
reflected in the magnetization per spin.  The horizontal parts of the plot
represent situations in which the domain wall remains unchanged, apart from
small fluctuations.  The steps correspond to events when critical
fluctuations appear, thereby causing small regions of the sample to reverse
their magnetization into the stable phase.  The average time between the
magnetization jumps naturally depends on the size of the sample observed.
For large regions it is expected to scale as the inverse of the length of
the domain wall. Similarly, the size of an individual magnetization jump is
proportional to the fraction of the observation area that is switched during
that jump.  Therefore, the magnetization curve should appear smooth for
observation areas much larger than a typical island size. 

It is interesting to note that similar jumps in the magnetization have been
observed experimentally in some thin magnetic films.\cite{Gadetsky} Although
in the present case the typical size of these Barkhausen jumps is about ten
times smaller than in that study, it is possible that the effect could be
observed using similar magneto-optical techniques.

While traversing the sample, the domain wall encounters barriers of
different heights, and the total time needed to travel a certain distance is
the sum of contributions similar to Eq.~(\ref{eq:nucl}), but with different
values of $\Xi$. As our simulations show, the residence time in different
metastable domain-wall configurations can be very different, suggesting that
there is a broad distribution of free-energy barriers. Since we cannot
determine that distribution, we assume it to be uniform between zero and an
upper cutoff $\Delta$.  This and Eq.~(\ref{eq:nucl}) lead us to an
approximate formula for the time elapsed during the domain-wall propagation
through a given sample (we only keep the dominant terms),
\begin{equation}
t_{\rm sw} \approx {A \over \mu H} + 
      {B\over (\mu H)^2} \exp\left({\Delta \over k_{\rm B} T \> \mu H}\right) \ .
\label{eq:fit}
\end{equation}
Here we have added the first term, which corresponds to the time elapsed
during the ``free'' domain-wall propagation, when the velocity is
proportional to the driving magnetic field $H$.  This term is usually
negligible, except in systems with coverage close to two. The parameters
$A$, $B$ and $\Delta$ are fit to the simulation data as described below.

\begin{figure}[t]
\vspace*{6.6in} 
\includegraphics{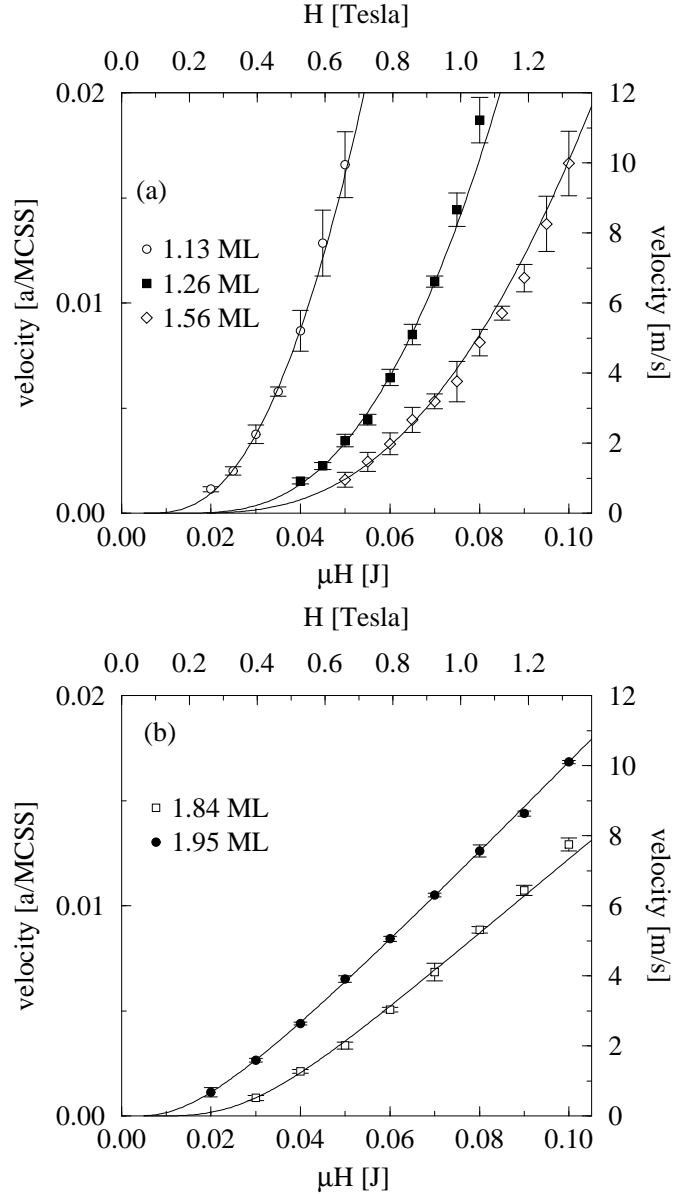} 
\caption[]{ Average domain-wall velocity as a function of the driving magnetic
field for several different coverages at $T=184$~K. As the coverage
increases towards $\approx 1.56$ ML, the field near which the velocity
practically vanishes increases (a). For coverages closer to 2 ML, it
decreases again with increasing coverage (b). 
\label{fig:velo}}
\end{figure}

The effective domain-wall velocity $v_{\rm eff} = D/t_{\rm sw}$ (with $D$
standing for the traveled distance, defined as the switched volume divided by
the width of the lattice), obtained from the averaged switching times, is
shown in Fig.~\ref{fig:velo} as a function of the driving field for several
different coverages of the sesquilayer systems. A qualitatively similar
domain-wall velocity dependence on driving field was found in experiments on
magnetic films,\cite{Sayko,Pokhil,Kirilyuk} in which the domain-wall motion
is also believed to be controlled by thermal fluctuations.

What is shown in Fig.~\ref{fig:velo} is an analogue of a depinning
transition rounded by a finite temperature. It is to be stressed that in the
present case there is no sharp transition, even in an arbitrarily weak
field. Strictly speaking, the average domain-wall velocity is always
nonzero, but it quickly approaches zero on a scale of field strengths which,
in agreement with the experimental findings, depend nonmonotonically on the
coverage.

We measured the switching times for several sesquilayer samples with various
coverages as functions of the magnetic field. Using Eq.~(\ref{eq:fit}), we
estimated the fitting parameters for each coverage.  The resulting values
are summarized in Table I. We note that the behavior of the fitting
parameters as functions of the coverage is quite reasonable.  Also, the
effective velocity of the ``free domain-wall motion'' $D \mu H/A$ is similar to
what has been measured in pure Ising systems.\cite{Ramos}

We want to point out the difference between our Eqs.~(\ref{eq:nucl},
\ref{eq:fit}) and an approach commonly used in the experimental
literature,\cite{Sayko,Pokhil} in which one supposes that the free-energy
barriers decrease {\em linearly} with $H$. That would be appropriate only if
the size of the critical fluctuations were fixed. The
scaling argument leading to Eq.~(\ref{eq:nucl}) implies that that is not so
in the present case.  However, it is extremely difficult to distinguish
numerically between linear and inverse linear dependencies if one only has
data from a relatively narrow range of field strengths. That is usually the
case in experiments, and we face the same problem here. However, attempting
to fit our simulation results using a linear field dependence, we obtain
fitting parameters which behave rather erratically as functions of the
coverage. We therefore believe that the ``$1/H$ model'' used here is more
appropriate in the range of $H$ studied here, also from the numerical point
of view.

Another question concerning the functional form of the velocity-field
relation used here, is whether it may be extrapolated to fields
sufficiently weak that the velocity vanishes on the experimental time scale.
We know that in a relatively strong field the size of the critical
fluctuation $l$ varies as $l \sim 1/H$. But is this still true when $l$
exceeds the length scales characteristic of the island structure of the
film? Or, equivalently, is the disorder strong enough to provide for a kind
of ``weak pinning'' on large length scales, with groups of islands acting as
``collective'' pinning centers?  Both qualitative arguments and exploratory
simulations of simplified configurations indicate that this is the case. 
However, the coercivities extracted from our simulations approximately
correspond to magnetic fields in which $l$ approaches the island size.  We
cannot exclude that in even weaker fields the relation of the velocity to
$H$ is characterized by parameters different from those obtained in the
relatively strong-field region to which our simulations are confined.
Without being able to perform simulations in extremely weak fields, the only
way to judge the appropriateness of our extrapolation is the comparison of
our final results with the experimental data.  The good, semiquantitative
agreement is an indication that our expectation is, indeed, correct.

Having an approximation formula for the domain-wall velocity $v_{\rm
eff}(H)$ as a function of the field, we can estimate the coercivity $H_c$.
Since coercivity is usually measured in a time-dependent field, we have to
translate our static measurements into such a situation.  We restrict
ourselves to relatively slowly varying and weak fields in which the
domain-wall motion is the main magnetization switching mechanism.  Suppose
that the system is subjected to a magnetic field $H_0 \sin(\omega t)$. The
average domain-wall displacement at time $t$ is then $\int_0^t v_{\rm
eff}(H_0 \sin(\omega t'))  dt'$.  Denote by $L_{\rm S}$ the typical distance
between seeds of the stable phase. Then, the coercivity can be estimated as
the value of the driving field reached when the domain walls have traveled
that distance,
\begin{equation}
{ a \over t_{\rm MC} \ L_{\rm S} \omega} 
     \int_0^{\arcsin(H_c/H_0)} v_{\rm eff}(H_0 \sin(\phi)) d\phi
\approx 1 \ .
\label{eq:coer}
\end{equation}
The velocity $v_{\rm eff}$ is given in dimensionless units, and the conversion
factor to ${\rm ms}^{-1}$, $a / t_{\rm MC}$, is shown explicitly. 
Note that this equation is determined up to a factor of order unity. Since
$v_{\rm eff}$ depends strongly on $H$, this does not affect the
estimated coercivity significantly.  It is the order of magnitude of the
combination $ a/(t_{\rm MC} L_{\rm S} \omega) $ and the value of the barrier
$\Delta$ which determine the coercivity.

There are two quantities in our model which we cannot determine precisely,
namely the typical distance $L_{\rm S}$ between the stable-domain seeds, and
the factor $t_{\rm MC}$ which relates the Monte Carlo and the physical time
scales. They always enter our formulas as a product, so there is a single
undetermined parameter in the theory which can be regarded as a fitting
parameter fixing the coercivity scale.  We use the following values for
$t_{\rm MC}$, $L_{\rm S}$, and $\omega$:  $t_{\rm MC} = 10^{-12}$ s, $L_{\rm
S} = 10^{-5}$ m, and $\omega = 2 \pi$ s$^{-1}$. While these are
``reasonable'' values, they were simply chosen such that the coercivity
simulated at $0.8 T_c$ (184~K) exhibits a maximum value (as a function of
coverage) which roughly corresponds to the experimental
observation.\cite{Skomski96} 

We used $H_0 = 0.5$~T for the value of the driving-field amplitude, 
which is about the field strengths used in 
experiments.\cite{Sander96a,Sander96b,Suen} 
When the coercivity $H_c$ is much smaller than $H_0$,
it depends only weakly on the amplitude and frequency of the driving 
field in the combination $\omega H_0$. 
On the other hand, one observes much stronger 
dependence on the amplitude when $H_c$ and $H_0$ are comparable.
In the extreme case when $H_0$ is less than the coercivity,
there is no solution to our Eq.~(\ref{eq:coer}). 
This situation corresponds to
the experimentally observed collapse of the hysteresis loop.\cite{Suen}

\begin{figure}[t]
\vspace*{2.8in} 
\includegraphics{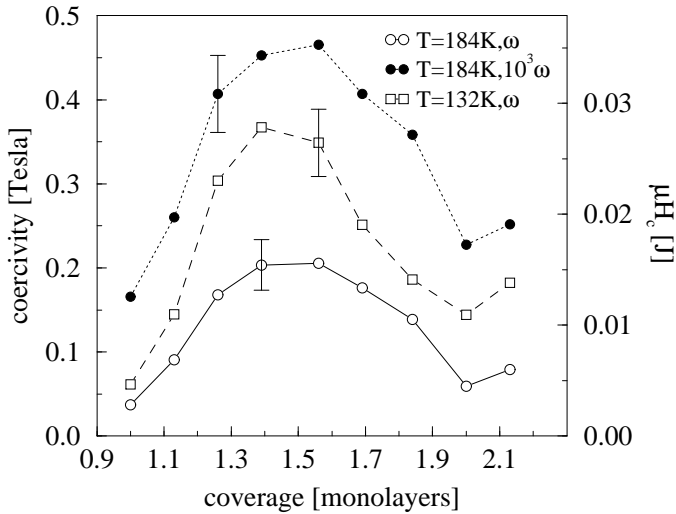} 
\caption[]{ Coercivity, $H_c$, of the iron sesquilayer as a function of
coverage. Different curves represent measurements at different temperatures
and frequencies.  The lower curve is to be compared with Fig.~3 of
Ref.~\protect{\onlinecite{Skomski96}}.  Representative error bars are shown
to give a rough idea of the statistical uncertainty due to errors in the
fitting parameters. 
\label{fig:coer}} 
\end{figure}

In Fig.~\ref{fig:coer} we plot the coercivity $H_c$ as a function of the
sesquilayer coverage.  The coercivity first steeply increases to a maximum
at a coverage of about 1.4. Then it decreases as the coverage approaches a
full double layer. This is very similar to the experimentally observed
behavior of the coercivity. As the temperature decreases, the coercivity
rapidly increases. Sander et~al. reported a coercivity higher than 0.3~T at
140~K.\cite{Sander96a} Our simulation in that temperature range nicely
reproduces that observation.  To illustrate the effect of the driving-field
frequency $\omega$, we also show a coercivity curve (filled circles)
predicted for a measurement at a frequency 1000 times higher than the one
shown by open circles.

It is expected that another coercivity maximum would be observed for
coverages between 2 and 3, although it should be much less
pronounced.\cite{Sander96a} We do not have suitable experimental data to
create lattice structures for this coverage region, but simulations on
structures which simply repeat the morphology of the thinner films do
suggest the existence of the second maximum.

\section{Conclusion}

In conclusion, the good agreement between the shapes and the dependencies on
$T$ and $\omega$ of the simulated and experimental plots of coercivity 
vs.\ coverage suggests that a simple kinetic Ising model can describe the basic
processes involved in domain-wall motion at nonzero temperatures in highly
anisotropic films.  Our simulations corroborate that the coercivity maximum
observed for iron sesquilayers on W(110) is caused by domain-wall pinning at
the islands of the second monolayer.  Although the pinning mechanism here is
in principle similar to the one proposed in the micromagnetic model of
Ref.\onlinecite{Sander96a}, the crucial difference is in the role of thermal
fluctuations, which the present model takes into account.  Thermal
fluctuations allow the domain wall to overcome obstacles created by the
second-monolayer islands by creating critical fluctuations of the
domain-wall shape. The domain-wall motion is intermittent with long periods
of quiescence punctuated by nucleation events which are followed by rapid
forward jumps of the wall. Without any further assumptions, our model also
reproduces an increase in the coercivity at lower temperatures.  It is to be
stressed that there is only one unknown parameter in the model. Once it is
fixed, simulations at different temperatures and film coverages can be used
to predict the coercivity. 

\begin{figure}[t]
\vspace*{5.3in} 
\includegraphics{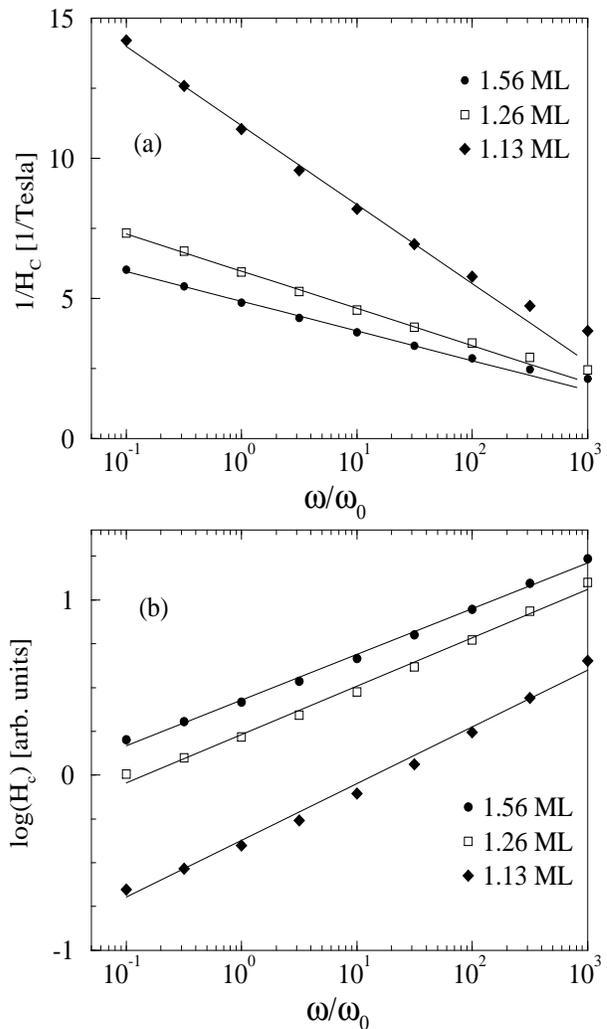} 
\caption[]{ The proposed model predicts that the coercivity depends on the
time scale of the measurement. Approximately, the inverse of the coercivity
is linear in the logarithm of the magnetic field sweep rate (frequency), as
shown in panel (a). The same data is shown also as $\log(H_c)$ vs.
$\log(\omega)$ in panel (b). Note that the plots are nearly linear in both
cases.  Points represent calculations based on our simulation data, and
straight lines are merely guides to the eye.  With $\omega_0 = 2\pi$, the
shown frequency range roughly corresponds to the one in
Ref.~\protect{\onlinecite{Suen}}.
\label{fig:freq}} 
\end{figure}

Another interesting point is that the model also predicts how the coercivity
depends on the sweep rate (or frequency) of the magnetic field in which the
measurement is done.  
By integrating Eq.~(\ref{eq:coer}) in the low-frequency limit one can show
that the inverse coercivity depends approximately linearly on the logarithm
of the sweep rate (frequency).  Figure~\ref{fig:freq}(a) shows that such an
approximation is quite reasonable.  However, the same data plotted on a
$\log(H_c)$ vs.\ $\log(\omega)$ scale in Fig.~\ref{fig:freq}(b) also provide
a quite good linear plot.  Recent experiments by Suen et~al.\cite{Suen}
show that the logarithm of the hysteresis loop area (which, because of the
shape of the loop, can be replaced by the coercivity in this case) 
increases linearly with $\log(\omega)$ over several decades.  Thus, our
observation is in good agreement with experimental findings, but it also
demonstrates that it can be difficult to draw conclusions about the
functional form of the frequency dependence, even with data spanning several
frequency decades. 

Finally, we note that the proposed model is interesting not only from the
point of view of its application to sesquilayer iron films. It describes the
basic physics of thermally activated domain-wall motion, which is common to
many experimental systems (for example the perpendicularly magnetized
ultrathin films studied in
 Refs.~[\onlinecite{Gadetsky,Sayko,Pokhil,Kirilyuk,He,Qiu}]) for which,
when suitably modified, the kinetic Ising model could provide a tool for a better
understanding of many effects. Ising-like models with a higher number
of states can be used to study rough films with a four-fold 
anisotropy\cite{Moschel} or can serve as an alternative to the
spin-1 mean-field-type description of the influence of the island structure
on the coercivity.\cite{Zangwill}
The dynamics of the model is also closely related
to more general topics such as interface propagation in 
disordered media\cite{Nowak} 
and depinning transitions in the presence of thermal fluctuations or
hysteretic creep of elastic manifolds.\cite{Scheidl}

The authors wish to thank A. Zangwill for useful discussions.
This research was supported by NSF Grants No. DMR-9315969 and DMR-9520325,
FSU-MARTECH, FSU-SCRI (DOE Contract No. DE-FC05-85ER25000).

\begin{table}
\begin{tabular}{|c|c|c|c|} 
coverage & \ $A/D\ [J]$  & \ $B/D\ [J^2]$ & \ $\Delta/k_{\rm B} T\ [J]$ \\ 
\hline
0.95 &\ $-$      &\ 0.08(1) \   &\    0.007(7) \ \\ \hline
1.13 &\ $-$      &\ 0.08(1) \   &\    0.033(2) \ \\ \hline  
1.26 &\ $-$      &\ 0.16(1) \   &\    0.075(5) \ \\ \hline 
1.39 &\ $-$      &\ 0.15(3) \   &\    0.100(15)\ \\ \hline
1.56 &\ $-$      &\ 0.22(2) \   &\    0.096(3) \ \\ \hline
1.69 &\ 5.1(5) \ &\ 0.10(1) \   &\    0.090(8) \ \\ \hline
1.84 &\ 6.2(3) \ &\ 0.12(2) \   &\    0.059(5) \ \\ \hline
1.95 &\ 4.5(4) \ &\ 0.12(2) \   &\    0.015(4) \ \\ \hline
2.13 &\ 5.1(5) \ &\ 0.09(2) \   &\    0.026(5) \  
\end{tabular}
\vspace{0.25 truecm}
\caption{ Fitting parameters of $v_{\rm eff} = D/t_{\rm sw}$ for several
coverages at $T=184$~K.  Because of insufficient data acuracy, only a
two-parameter fit was used for low coverages, where the term proportional to
$A$ is small compared to the exponential contribution in
Eq.~(\protect{\ref{eq:fit}}).  Numerical uncertainities were estimated by
processing two or more independent data sets.  
}
\end{table}


\begin{references}


\bibitem{Bethge95}
H. Bethge, D. Heuer, Ch. Jensen, K. Resh\" oft and U. K\" ohler,
Surf. Sci. {\bf 331-333}, 878 (1995).

\bibitem{Sander96a}
D. Sander, R. Skomski, C. Schmidthals, A. Enders and J. Kirschner,
Phys.\ Rev.\ Lett. {\bf 77}, 2566 (1996).

\bibitem{Sander96b}
D. Sander, A. Enders, R. Skomski and J. Kirschner,
IEEE Trans.\ Magn. {\bf 32}, 4570 (1996).

\bibitem{Skomski96}
R. Skomski, D. Sander, A. Enders and J. Kirschner,
IEEE Trans.\ Magn. {\bf 32}, 4567 (1996).

\bibitem{Elmers95}
H.\ J. Elmers, J. Hauschild, H. Fritzsche, G. Liu and U. Gradmann,
Phys.\ Rev.\ Lett. {\bf 75}, 2031 (1995).

\bibitem{Elmers94}
H.\ J. Elmers, J. Hauschild, H. H\" oche, U. Gradmann, 
H. Bethge, D. Heuer and U. K\" ohler,
Phys.\ Rev.\ Lett. {\bf 73}, 898 (1994).

\bibitem{Elmers90}
H.\ J. Elmers and U. Gradmann,
Appl.\ Phys.\ A {\bf 51}, 255 (1990).

\bibitem{Suen}
J. H. Suen and J. L. Erskine,
Phys.\ Rev.\ Lett. {\bf 78},  3567 (1997).

\bibitem{Rikvold}
P.~A.\ Rikvold, H.\ Tomita, S.\ Miyashita and S.~W.\ Sides,
Phys.\ Rev.\ E {\bf  49}, 5080 (1994);
P.~A.\ Rikvold and B.~M.\ Gorman, in Ann. Rev. Comp. Phys. I,
edited by D. Stauffer (World Scientific, Singapore, 1994), p. 149.

\bibitem{Richards}  H.~L.\ Richards et al.,
J.\ Magn.\ Magn.\  Mater. {\bf 150}, 37 (1995); 
J.\ Appl.\ Phys.\ {\bf 79}, 5479 (1996);
Phys.\ Rev.\ B {\bf 54}, 4113 (1996);
Phys.\ Rev.\ B {\bf 55}, 11521 (1997).

\bibitem{Geilo97}
P.~A.\ Rikvold, M.\ A. Novotny and M. Kolesik, preprint cond-mat/9705189.

\bibitem{Martin77}
Ph.~A. Martin, J.\ Stat.\ Phys. {\bf 16}, 149 (1977).

\bibitem{Novotny}
M.\ A. Novotny, Phys.\ Rev.\ Lett. {\bf 74}, 1 (1995);
Erratum {\bf 75}, 1424 (1995);
Comp. in Phys. {\bf 9}, 46 (1995).

\bibitem{Gadetsky}
S. Gadetsky and M. Mansuripur, 
J.\ Appl.\ Phys.\ {\bf 79}, 5667 (1996).

\bibitem{Sayko}
G.\ V. Sayko, A.\ K. Zvezdin, T.\ G. Pokhil, 
B.\ S. Vvedensky and E.\ N. Nikolaev,
IEEE Trans. Magn. {\bf 28}, 2931 (1992).

\bibitem{Pokhil}
T.\ G. Pokhil and E.\ N. Nikolaev, 
IEEE Trans. Magn. {\bf 29}, 2536 (1993).

\bibitem{Kirilyuk}
A. Kirilyuk, J. Ferr\' e and D. Renard, 
IEEE Trans. Magn. {\bf 29}, 2518 (1993).

\bibitem{Ramos}
R. Ramos, S.\ W. Sides, P.\ A. Rikvold, M.\ A. Novotny,  
in preparation.

\bibitem{He}
Y.-L. He and G.-C. Wang, Phys.\ Rev.\ Lett. {\bf 70}, 2336 (1993).

\bibitem{Qiu}
Z.\ Q. Qiu, J. Pearson and S.\ D. Bader,
Phys.\ Rev. B {\bf 49}, 8797 (1993).


\bibitem{Moschel}
A. Moschel, R.~A. Hyman, A. Zangwill and M.~D. Stiles,
Phys.\ Rev.\ Lett. {\bf 77}, 3653 (1996).

\bibitem{Zangwill}
C.~N. Luse and A. Zangwill,
J.\ Appl.\ Phys.\ {\bf 79}, 4942 (1996).

\bibitem{Nowak}
U. Nowak, J. Heimel, T. Kleinfeld and D. Weller,
Phys.\ Rev. B in press.


\bibitem{Scheidl}
S. Scheidl and V. Vinokur, 
Phys. Rev. Lett. {\bf 77}, 4768 (1996).

\end{references}
\end{document}